\documentclass[11pt]{article}
\usepackage[margin=1in]{geometry}
\usepackage{graphicx}
\usepackage{hyperref}
\usepackage{booktabs}
\usepackage{amsmath}
\usepackage{listings}
\usepackage{xcolor}

\title{Two-Path Status Verification for Outbound Enterprise Messaging Pipelines: Webhook and Scheduled Polling Fallback Architecture}
\author{Devam Gupta \\ Staff Engineer, Twilio \\ Technical Architect, Twilio for Salesforce}
\date{July 2026}

\begin{document}
\maketitle

\begin{abstract}
Outbound enterprise messaging pipelines face a fundamental reliability challenge: delivery status callbacks (webhooks) from messaging providers are subject to network failures, endpoint unavailability, and provider-side retry exhaustion, resulting in stale status records in the CRM system of record. A naive single-path architecture that relies exclusively on webhooks leaves a population of messages permanently in an intermediate state when callbacks fail. This paper presents a two-path status verification architecture, generalized from patterns observed in production CRM-native messaging systems built on multi-tenant platform-as-a-service infrastructure. The primary path uses a real-time webhook received by a REST endpoint, which publishes an internal event for asynchronous record update. The fallback path uses a configurable scheduled polling job that detects records still in transitional status after a configurable interval and queries the provider's status API directly to reconcile state. We describe the event-driven primary path, the scheduler-based fallback, deduplication via idempotent upsert, the sync failure detection mechanism, and the platform resource-limit considerations that shape each design decision.
\end{abstract}

\noindent\textbf{Keywords:} Webhook Reliability, Enterprise Messaging, Platform-as-a-Service, Event-Driven Architecture, Idempotency, Scheduled Jobs, Status Reconciliation, Multi-Tenancy.

\section{Introduction}

When an enterprise CRM system sends an outbound message through a third-party provider, the message transitions through a sequence of states managed by that provider: queued $\rightarrow$ sending $\rightarrow$ sent $\rightarrow$ delivered (or failed/undelivered). The CRM system requires accurate, up-to-date status on each message for agent visibility, reporting, and operational monitoring. The provider communicates these status transitions to the CRM via HTTP POST callbacks -- webhooks -- to a registered endpoint.

Webhook delivery is inherently unreliable. The receiving endpoint may be temporarily unavailable. Network timeouts may cause the provider to exhaust its retry policy before the endpoint recovers. A managed-platform REST endpoint is typically subject to concurrent-request and execution-time limits~\cite{sfgovlimits2024}, and may reject or drop requests under load. In each of these failure modes, the status callback is lost and the record in the CRM retains its last-known status -- typically an intermediate ``sent'' state -- indefinitely.

For an enterprise messaging platform where agents monitor delivery status to detect failed messages and trigger follow-up actions, a population of permanently-stale intermediate-status records represents both an operational and a data-integrity problem.

This paper is a companion to two earlier papers describing the same class of production architecture: a treatment of the overall CRM-native messaging data model and send/receive pipeline~\cite{gupta2026arch}, and a treatment of an opt-in/opt-out consent data model~\cite{gupta2026consent}. Where those papers describe a system's data model and its outbound and inbound paths at a general level, this paper focuses specifically on the status-reconciliation problem: how such an architecture can guarantee that a message's status in the CRM eventually converges to the truth known by the messaging provider, even when the primary delivery mechanism for that information -- the webhook -- fails.

The architecture described in this paper solves the reconciliation problem with two independent verification paths: a real-time webhook path that handles the normal case, and a scheduled polling path that detects and resolves the fallback case. The two paths are designed to converge on the same correct state through an idempotent-upsert mechanism, with no risk of inconsistency when both paths act on the same message.

\section{System Architecture Overview}

At a conceptual level, the messaging pipeline consists of five components:

\begin{enumerate}
\item \textbf{Webhook Endpoint} -- a public REST resource that receives HTTP POST callbacks from the messaging provider for both inbound messages and outbound status updates.
\item \textbf{Status Event Channel} -- an internal asynchronous event, published by the webhook endpoint, that decouples HTTP request handling from CRM record processing, consistent with the inbound decoupling pattern described in~\cite{gupta2026arch}.
\item \textbf{Status Event Handler} -- a subscriber that processes the event and upserts the corresponding message record.
\item \textbf{Polling Scheduler} -- a periodic job implementing the fallback polling path, which detects sync failures and triggers direct provider-API reconciliation.
\item \textbf{Monitoring Job} -- a periodic job that detects prolonged sync failures and alerts administrators.
\end{enumerate}

\section{Primary Path: Webhook to Event Channel to Record Upsert}

\subsection{Webhook Endpoint}

The webhook endpoint is a REST resource registered at a public URL and exposed as a public HTTPS endpoint. At a high level, the handler performs the following steps on each inbound request:

\begin{lstlisting}
function handleCallback(request):
    params = request.parameters
    signature = request.headers["X-Provider-Signature"]
    endpoint = "https://" + request.headers["Host"]
        + request.path
\end{lstlisting}

Before processing, the endpoint validates three security conditions: (1) the account identifier in the request matches the configured account in the CRM, (2) an authentication token is configured, and (3) a cryptographic signature header validates correctly against the request parameters using an HMAC-based signature scheme. Requests failing any validation receive a 400 or 403 response and are not processed further.

\subsection{Event Publication}

Upon successful validation, the webhook handler maps the HTTP parameters onto a status-event payload and publishes it synchronously to the internal event channel:

\begin{lstlisting}
event = StatusEvent(
    body = params["Body"],
    to = params["To"],
    from_ = params["From"],
    accountId = params["AccountId"],
    messageId = params["MessageId"],
    serviceId = params["ServiceId"],
    status = params["Status"],
    errorCode = params["ErrorCode"]
)

eventBus.publish(event)
response.statusCode = 202
\end{lstlisting}

The 202 Accepted response is returned to the messaging provider immediately after publishing the event -- before any subscriber has processed it. This is a deliberate design: webhook response latency is decoupled from CRM record-processing latency, mirroring the inbound webhook decoupling pattern described in~\cite{gupta2026arch}. The messaging provider receives an acknowledgment within its HTTP timeout window regardless of downstream processing time.

\subsection{Status Event Handler}

The status-event subscriber fires asynchronously after publication. The handler filters events by account and service identifiers to ensure only events from the configured account are processed -- important in multi-tenant deployments where multiple CRM organizations may share underlying infrastructure.

The handler builds a map of message records keyed by the provider's message identifier, applying a state machine to handle out-of-order event delivery:

\begin{lstlisting}
switch (existingStatus):
    case "sent":
        if newStatus not in ("queued", "sending"):
            existing.status = newStatus
    case "delivered":
        if newStatus not in ("queued", "sending", "sent"):
            existing.status = newStatus
    // additional terminal and intermediate states
\end{lstlisting}

This prevents status regression: a ``delivered'' message cannot be moved back to ``sent'' by a delayed callback that arrives out of sequence. The state machine encodes a valid forward-only progression: queued $\rightarrow$ sending $\rightarrow$ sent $\rightarrow$ delivered (terminal) / undelivered / failed (terminal).

The final write is an idempotent upsert keyed on the provider's message identifier as an external-id field, executed through a centralized data-access layer that enforces object- and field-level permissions on every write, consistent with the access-control approach described in~\cite{gupta2026arch}:

\begin{lstlisting}
dataAccessLayer.upsert(
    updatedRecords,
    externalIdField = "ProviderMessageId",
    enforcePermissions = true
)
\end{lstlisting}

If the write fails -- for example, due to a transient platform resource-limit exception -- the handler relies on the event channel's built-in retry mechanism, retrying a bounded number of times before logging the failure and discarding the event, providing a secondary layer of resilience within the primary path itself.

\section{Fallback Path: Scheduled Polling}

\subsection{Design Rationale}

The fallback path activates when the primary webhook path has failed to deliver status updates for a configurable interval. Rather than continuously polling all messages -- which would exhaust API-callout limits~\cite{sfgovlimits2024} -- the fallback path operates on a targeted population: message records whose status has not been updated within the expected window.

\subsection{Scheduler Architecture}

The polling scheduler implements a self-rescheduling pattern: each execution cancels its own scheduled trigger and re-schedules itself for a future time based on a configurable poll interval:

\begin{lstlisting}
function onScheduledExecution(context):
    cancelScheduledTrigger(context.triggerId)
    enqueueJob(ReconciliationJob(mode = "BY_DATE_RANGE"))
\end{lstlisting}

The self-abort-and-reschedule pattern is used because many managed-platform schedulers do not support sub-minute intervals natively -- the minimum granularity is often one minute. By scheduling the next execution dynamically based on a configurable polling-interval setting, the effective polling interval can be set to values such as 3 or 5 minutes while ensuring exactly one scheduled job exists at any time.

A companion routine is invoked at the end of each reconciliation run to maintain continuity -- if no job is currently scheduled, it schedules the next one, guarded by a check against existing scheduled jobs to prevent duplicate scheduling.

\subsection{Message Reconciliation}

The reconciliation job, when invoked in its date-range mode, queries message records that remain in transitional states (sent, queued, sending) beyond the polling window. For each such record, it calls the messaging provider's status API to retrieve the current status and updates the record directly -- bypassing the event-channel path entirely, since this is a reconciliation operation rather than a real-time event.

\section{Sync Failure Detection and Alerting}

\subsection{Monitoring Job}

A third periodic job runs hourly and monitors a last-successful-sync timestamp stored in the messaging configuration. This timestamp is updated by the status event handler each time a message status update is successfully processed.

The monitoring job compares the current time against the last recorded sync time. If the elapsed time exceeds the configured polling interval plus a grace window (empirically, on the order of 10-15 minutes), it concludes that the sync pipeline has stalled and sends an alert to the configured administrator. The grace window accounts for legitimate gaps in message activity -- the absence of a recent sync timestamp does not necessarily indicate a failure if no messages have recently been exchanged.

\subsection{Notification Suppression}

Administrators can suppress sync-failure notifications via a configuration flag, allowing temporary suppression during maintenance windows without requiring code changes or scheduler modifications.

\section{Deduplication and Idempotency}

Both paths may process the same message status transition. The idempotent upsert on the provider's message identifier ensures that concurrent or duplicate processing produces a consistent final state rather than duplicate records. The state machine in the event handler further ensures that if the webhook path has already advanced a message to ``delivered'', a subsequent polling reconciliation that encounters a delayed ``sent'' callback does not revert the record.

This convergence property -- that both paths, applied in any order, produce the same final state -- is the key correctness guarantee of the two-path architecture. It allows the two paths to operate independently without coordination, with the upsert mechanism providing the synchronization. This convergence-by-idempotency approach extends the composite-key idempotent-upsert pattern used for opt-out processing in the consent architecture described in~\cite{gupta2026consent} to the status-tracking domain.

\section{Platform Resource-Limit Considerations}

Managed multi-tenant platforms commonly enforce a bounded number of external HTTP callouts per execution context~\cite{sfgovlimits2024}. The polling fallback path batches message reconciliation within an asynchronous job framework~\cite{sfqueueable2024}, which provides a fresh resource-limit context per execution. This allows the polling job to process large backlogs of stale messages across multiple chained job executions without hitting callout limits, using the same self-chaining pattern applied to bulk sends in~\cite{gupta2026arch}.

The event-channel path is largely exempt from the synchronous resource limits that govern the webhook request itself -- event subscribers typically execute in their own transaction context, with limits reset for each invocation. This architectural property is part of why an internal event bus is an appropriate intermediary for high-volume inbound webhook processing on managed multi-tenant platforms~\cite{sfplatformevents2024}.

The scheduled-job infrastructure relies on querying platform job metadata to check for existing scheduled jobs. Such queries are typically against system metadata rather than application data, and do not count against per-transaction data-query limits, making the duplicate-schedule guard inexpensive to call frequently.

\section{Discussion}

The two-path architecture described here reflects a general pattern applicable to any system where a real-time push-notification mechanism (webhooks, push events, server-sent events) is the primary delivery channel but cannot be relied upon exclusively. The pattern has three components: a primary push path for low-latency normal-case processing, a scheduled pull path for fallback reconciliation, and a convergence mechanism (idempotent upsert with a state machine) that ensures both paths produce consistent results regardless of execution order.

The event-channel intermediary in the primary path is significant beyond simple decoupling. By publishing the webhook payload as an internal event and returning a 202 response immediately, the system transforms a synchronous HTTP dependency into an asynchronous processing pipeline. The messaging provider receives acknowledgment within its timeout window even if the CRM's write layer is experiencing contention. This prevents webhook retry storms -- where a slow CRM response causes the provider to retry aggressively, amplifying load -- and ensures that the 202 response accurately reflects receipt of the event, not completion of processing.

The self-rescheduling scheduler pattern addresses a gap common to managed-platform scheduling infrastructure: such platforms often support cron-based scheduling at minute granularity but do not support continuous or sub-minute polling natively. The pattern of aborting and rescheduling within each execution achieves configurable polling intervals that are not limited to a scheduler's fixed cron-expression granularity.

Read together with~\cite{gupta2026arch} and~\cite{gupta2026consent}, this paper completes a three-part account of a class of production architecture: the general CRM-native data model and send/receive pipeline, the consent and compliance data model, and the status-reconciliation mechanism that keeps all of the above consistent under real-world network and platform failure conditions.

\section{Conclusion}

This paper has presented a two-path status verification architecture for enterprise CRM-native outbound messaging, generalized from patterns observed in production managed-package deployments on multi-tenant platform-as-a-service infrastructure. The primary webhook path leverages an internal event channel to decouple HTTP endpoint response latency from CRM record processing, with a state machine that handles out-of-order status-callback delivery. The fallback polling path uses a self-rescheduling scheduler to detect and reconcile messages that the webhook path failed to update, with configurable intervals and administrator alerting when the sync pipeline stalls. The convergence of both paths through idempotent upsert on a stable external identifier ensures consistency regardless of which path processes a given status update. These patterns are applicable to any enterprise integration where high-reliability status tracking is required against a push-notification provider that cannot guarantee delivery.

\end{document}